\documentstyle[osa,manuscript,psfig]{revtex}

\begin{document}

\titlepage
\title{Invariant mass dependence of 
$\Lambda$ polarization in $pp\to p\Lambda K^+\pi^+\pi^-\pi^+\pi^-$}
\author {Liang Zuo-tang and Li Tie-shi}
\address{Department of physics, Shandong University, Jinan 
Shandong 250100, China}
\maketitle

\begin{abstract}                % DON'T CHANGE THIS LINE
We show that there is a correlation between 
the invariant mass 
of the produced $\Lambda K^+$, 
$\Lambda K^+\pi^+\pi^-$ or 
$\Lambda K^+ \pi^+\pi^-\pi^+\pi^-$ system 
in the exclusive reaction 
$pp\to p\Lambda K^+\pi^+\pi^-\pi^+\pi^-$ 
and the longitudinal or transverse momentum of $\Lambda$. 
Together with the longitudinal 
and transverse momentum dependence 
of $\Lambda$ polarization observed  
in inclusive reactions, 
such a correlation implies 
a dependence of $\Lambda$ polarization 
on these invariant masses. 
The qualitative features of this dependence 
are consistent with 
the recent observation by E766 collaboration at BNL. 
A quantitative estimation has been made using an 
event generator for $pp$ collisions. 
A detailed comparison with the data is made. 
\end{abstract}     

%\narrowtext
%\twocolumn
\newpage

Since the discovery \cite{Les75} of  
hyperon polarization ($P_H$) in 
inclusive production processes 
at high energies, 
there has been constant interest in studying 
the origin of this effect, 
both experimentally \cite{Heller96} 
and theoretically [\ref{Kane78}-\ref{LB97}].  
It is now an established experimental fact that, 
in high energy hadron-hadron or hadron-nucleus collisions, 
the produced hyperons are polarized transversely 
to the production plane, 
although neither the projectiles 
nor the targets are polarized before the collisions. 
A large number of experiments on the inclusive production processes 
show that the polarization is significantly different from zero 
in the fragmentation regions 
for moderately large transverse momenta. 
The magnitude of the polarization increase with 
increasing $x_F$ and $p_\perp$, 
where $x_F\equiv 2p_\parallel /\sqrt{s}$ is the Feynman-$x$, 
$s$ is the total center of mass energy squared, 
$p_\|$ and $p_\perp$ are respectively 
the longitudinal and transverse component of 
the momentum of the produced hyperon.

Recently, further progresses 
have been achieved in experimental studies.  
An interesting program has been 
established by E766 Collaboration at 
Brookheaven National Laboratory (BNL) 
to study $\Lambda$ polarization 
$P_\Lambda$ in different 
specific reaction channels.  
The first experiment  
has already been carried out \cite{Fel96} 
on $P_\Lambda$ in $pp\to p\Lambda K^+\pi^+\pi^-\pi^+\pi^-$. 
They found in particular 
the following interesting phenomenon:
$P_\Lambda$ in this channel 
depends not only on $x_F$ and $p_\perp $ 
of the produced $\Lambda$ 
(as observed earlier in inclusive experiments), 
but also on the invariant masses 
$M_{\Lambda K}$, $M_{\Lambda K\pi\pi}$, 
and $M_{\Lambda K\pi\pi\pi\pi}$, 
of the particle systems $\Lambda K^+$, 
$\Lambda K^+\pi^+\pi^-$ 
and $\Lambda K^+\pi^+\pi^-\pi^+\pi^-$ respectively. 
The magnitude of $P_\Lambda$ increases with the increasing 
of these invariant masses $M_{\Lambda K}$, $M_{\Lambda K\pi\pi}$, 
and $M_{\Lambda K\pi\pi\pi\pi}$.

We recall that, compared with those on inclusive processes,  
the experiments on exclusive processes 
have the advantage to  
study the dependence on 
kinematic variables other than $x_F$ or $p_\perp$.
It offers the opportunity 
to investigate different correlations 
which usually provide us with information 
that can not be obtained in inclusive experiments. 
Such information gives often deeper insight into 
the physics behind the data. 
Hence, it is conceivable that the above mentioned 
new experimental finding provide further important tests 
of the different models and give 
us some new clue in the searching of the origin of 
$\Lambda$ polarization in high energy 
hadron-hadron collisions.
However, kinematic analysis readily shows that, 
due to energy-momentum conservation, 
there has to be a correlation between the above-mentioned 
invariant mass $M_{\Lambda K}$, $M_{\Lambda K\pi\pi}$, 
or $M_{\Lambda K\pi\pi\pi\pi}$
and $x_F$ or $p_\perp$ of $\Lambda$. 
On the average, both $x_F$ and $p_\perp$ increase 
with the increasing of these invariant masses 
$M_{\Lambda K}$, $M_{\Lambda K\pi\pi}$, 
or $M_{\Lambda K\pi\pi\pi\pi}$.
Together with the $x_F$ and $p_\perp$ 
dependences of $P_\Lambda$ 
observed earlier in the inclusive experiments, 
such correlations lead already to some increase 
of $|P_\Lambda|$ with the increasing of 
$M_{\Lambda K}$, $M_{\Lambda K\pi\pi}$, 
or $M_{\Lambda K\pi\pi\pi\pi}$. 
Hence, we are led naturally to the following questions:
How large are these correlations?
Can they already account for the observed 
invariant mass dependence of $P_\Lambda$?
Do the observed increase of $P_\Lambda$ with increasing 
$M_{\Lambda K}$, $M_{\Lambda K\pi\pi}$, 
or $M_{\Lambda K\pi\pi\pi\pi}$
give us some further tests of the 
different models or they are just different manifestations 
of the $x_F$ and $p_\perp$ dependences 
observed earlier in inclusive experiments?
These are questions which we would like to investigate 
in this note.

We recall that the invariant mass of 
$\Lambda K^+$-system is defined as 
the total energy in their center of 
mass (c.m.) system, i.e., 
\begin{equation}
M_{\Lambda K}=\sqrt{m_\Lambda^2+p^{*2}}+\sqrt{m_K^2+p^{*2}},
\end{equation}
where $m_K$ and $m_\Lambda$ are their masses, $p^*$ 
is their momentum in the c.m. frame of this particle system. 
It is obvious that, $M_{\Lambda K}$ 
increases with increasing $p^*$.
Large value of $p^*$ implies that the 
difference between the momentum of $\Lambda$ 
and that of $K^+$ is large, also 
in the c.m. frame of the colliding $pp$-system.  
Hence, we expect 
large difference between $x_F$ (or $p_\perp$) 
of $\Lambda$ and that of $K^+$ for large $M_{\Lambda K}$. 
On the other hand, 
if $x_F$ is very large (say, larger than $0.5$),  
the longitudinal momentum of $\Lambda$ is very large. 
In this case, $\Lambda$ carries already a very large 
part of the momentum of the whole system. 
According to energy-momentum conservation, 
the momentum of $K^+$ cannot be the same as that for $\Lambda$ 
since the sum of them cannot exceed $\sqrt{s}/2$. 
The longitudinal momentum for $K^+$ has to be much smaller.
This implies a large difference 
between them thus a large $M_{\Lambda K}$.
Hence, we expect that $M_{\Lambda K}$ 
increases with increasing $x_F$ for large $x_F$.
Similarly, if $p_\perp$ of $\Lambda$ is large, 
there should be a large probability that $p_\perp$ 
of $K^+$ is large and in the opposite direction 
to guarantee the transverse momentum conservation. 
This leads to a large difference in transverse momenta 
for $\Lambda$ and $K^+$ thus also a large $M_{\Lambda K}$.
Hence, we expect $M_{\Lambda K}$ increases also 
with increasing $p_\perp$.
We are naturally led to the following questions: 
How fast does $M_{\Lambda K}$ increase with 
increasing $x_F$ or $p_\perp$?
Are there similar correlations between 
$M_{\Lambda K\pi\pi}$ (or $M_{\Lambda K\pi\pi\pi\pi}$) 
and $x_F$ or $p_\perp$ of $\Lambda$?   
Apparently, the answers to these questions 
are determined by the momentum distributions 
of the particles produced in the collision processes 
and these distributions can be 
obtained in unpolarized reactions.

In order to study these questions quantitatively, 
we used a Monte-Carlo event generator PYTHIA \cite{Pythia}.
We recall that PYTHIA is an event generator 
for unpolarized high energy hadronic reactions 
based on Lund fragmentation model \cite{And83}.
It describes most (if not all) 
of the different features 
of the data for particle production in 
unpolarized reactions. 
We therefore expect that we should be able to 
obtain a reasonable description of 
the correlations between $x_F$ or $p_\perp$ 
and the invariant masses mentioned above. 
We generated 100,000 
$pp\to p\Lambda K^+\pi^+\pi^-\pi^+\pi^-$ events using PYTHIA. 
From these events, we calculated 
the average values of $x_F$ and those of $p_\perp$ 
for different $M_{\Lambda K}$, 
$M_{\Lambda K\pi\pi}$ or $M_{\Lambda K\pi\pi\pi\pi}$ bins.
The obtained results are shown in Figs.1 and 2. 
From Figs.1(a) and 2(a), we see clearly that 
both $\langle x_F\rangle $ and $\langle p_\perp\rangle $
increase with increasing $M_{\Lambda K}$. 
This is consistent with our qualitative expectations.
From Figs.1(b),1(c),2(b) and 2(c), we see also similar 
correlations between 
$M_{\Lambda K\pi \pi}$ or $M_{\Lambda K\pi\pi\pi\pi}$ 
and $x_F$ or $p_\perp$. 
The average values $\langle x_F\rangle $ 
and $\langle p_\perp\rangle $ 
increase also with increasing $M_{\Lambda K\pi\pi}$ or 
$M_{\Lambda K\pi\pi\pi\pi}$. 
But, we also see that the magnitudes 
of these correlations are smaller
than that between $x_F$ or $p_\perp $ 
and $M_{\Lambda K}$.

Since the magnitude of $\Lambda$ polarization $P_\Lambda$ 
increases with increasing $x_F$ or increasing $p_\perp $, 
we expect that $|P_\Lambda|$ has to increase also with 
$M_{\Lambda K}$, $M_{\Lambda K\pi\pi}$, 
or $M_{\Lambda K\pi\pi\pi\pi}$ 
because of the above mentioned correlations. 
This qualitative feature is in agreement with the data\cite{Fel96}. 
To see whether this effect explains also quantitatively 
the observed dependences of $P_\Lambda$ on 
$M_{\Lambda K\pi\pi}$, $M_{\Lambda K\pi\pi}$ 
or $M_{\Lambda K\pi\pi\pi\pi}$, 
we did the following calculations.
We assume that $P_\Lambda$ is 
determined completely by  
the $x_F$ and $p_\perp$ of $\Lambda$ 
and use the $x_F$ and $p_\perp$ dependences 
of $P_\Lambda$ obtained in experiments on 
inclusive reactions as input to calculate 
$P_\Lambda$ as a function of 
$M_{\Lambda K}$,  that of $M_{\Lambda K\pi\pi}$    
and that of  $M_{\Lambda K\pi\pi\pi\pi}$ respectively. 
The results for $P_\Lambda$ 
as a function of $x_F$ and $p_\perp$ obtained earlier in 
inclusive experiments can 
be parametrized as \cite{Lun89}
\begin{equation}
P_{\Lambda}(x_F,p_\perp)=1.5(c_1x_F+c_2x_F^3)(1-e^{c_3p_\perp^2})
\end{equation}
where $c_1=-0.268\pm 0.003$,  
$c_2=-0.338\pm 0.015$,  and
$c_3=-4.5\pm 0.6$,  
are constants determined by the data.
Using this parameterization for 
$P_\Lambda (x_F,p_\perp)$, 
we obtain $P_\Lambda$ as functions of 
$M_{\Lambda K}$, $M_{\Lambda K\pi\pi}$ and 
$M_{\Lambda K\pi\pi\pi\pi}$ shown in Fig.3.

From Fig.3,  we see that, 
$|P_\Lambda|$ increases indeed with increasing 
$M_{\Lambda K}$, $M_{\Lambda K\pi\pi}$ 
or $M_{\Lambda K\pi\pi\pi\pi}$. 
This qualitative feature agrees 
with the data\cite{Fel96}.
We see also that,
the $M_{\Lambda K}$ dependence of 
$P_\Lambda$ agrees with the data even quantitatively.\cite{foot} 
This shows that the above mentioned kinematic effect 
together with the early observed $x_F$ and $p_\perp$ dependences 
of $P_\Lambda$ can already explain this $M_{\Lambda K}$ dependence.   
The qualitative features of $P_\Lambda$ 
as a function of $M_{\Lambda K \pi\pi}$ and that 
of $M_{\Lambda K \pi\pi\pi\pi}$ agree also with the data. 
But, quantitatively, the increase of $|P_\Lambda|$ 
is too slow and is not enough to 
account for the observed $M_{\Lambda K\pi\pi}$ 
or $M_{\Lambda K\pi\pi\pi\pi}$ dependence. 
In particular, $|P_\Lambda|$ seems too large 
near the threshold of 
$M_{\Lambda K\pi\pi}$ or $M_{\Lambda K\pi\pi\pi\pi}$.
This means that other dynamic effects have to be 
introduced to account for this effect.  
This also implies that these dependences 
should be other independent tests for 
the different theoretical models [\ref{Bo79}-\ref{LB97}].

In summary, we made a kinematic analysis 
of the dependence of $P_\Lambda$ in 
$pp\to p\Lambda K^+\pi^+\pi^-\pi^+\pi^-$ on 
the invariant masses of the produced 
$\Lambda K^+$, $\Lambda K^+\pi^+\pi^-$ and 
$\Lambda K^+\pi^+\pi^-\pi^+\pi^-$ systems. 
We showed that there is a correlation between 
these invariant masses and $x_F$ or $p_\perp$ of $\Lambda$. 
This correlation gives already a 
reasonable explanation of the 
increase of $|P_\Lambda|$ with increasing $M_{\Lambda K}$ 
but not that with increasing $M_{\Lambda K\pi\pi}$ 
or $M_{\Lambda K\pi\pi\pi\pi}$.

We thank Xie Qu-bing, Wang Qun, Si Zong-guo 
for helpful discussions. This work is supported 
in part by the National Natural Science Foundation (NSFC) 
and the Education Ministry of China.

\newpage

\begin{figure}
\psfig{file=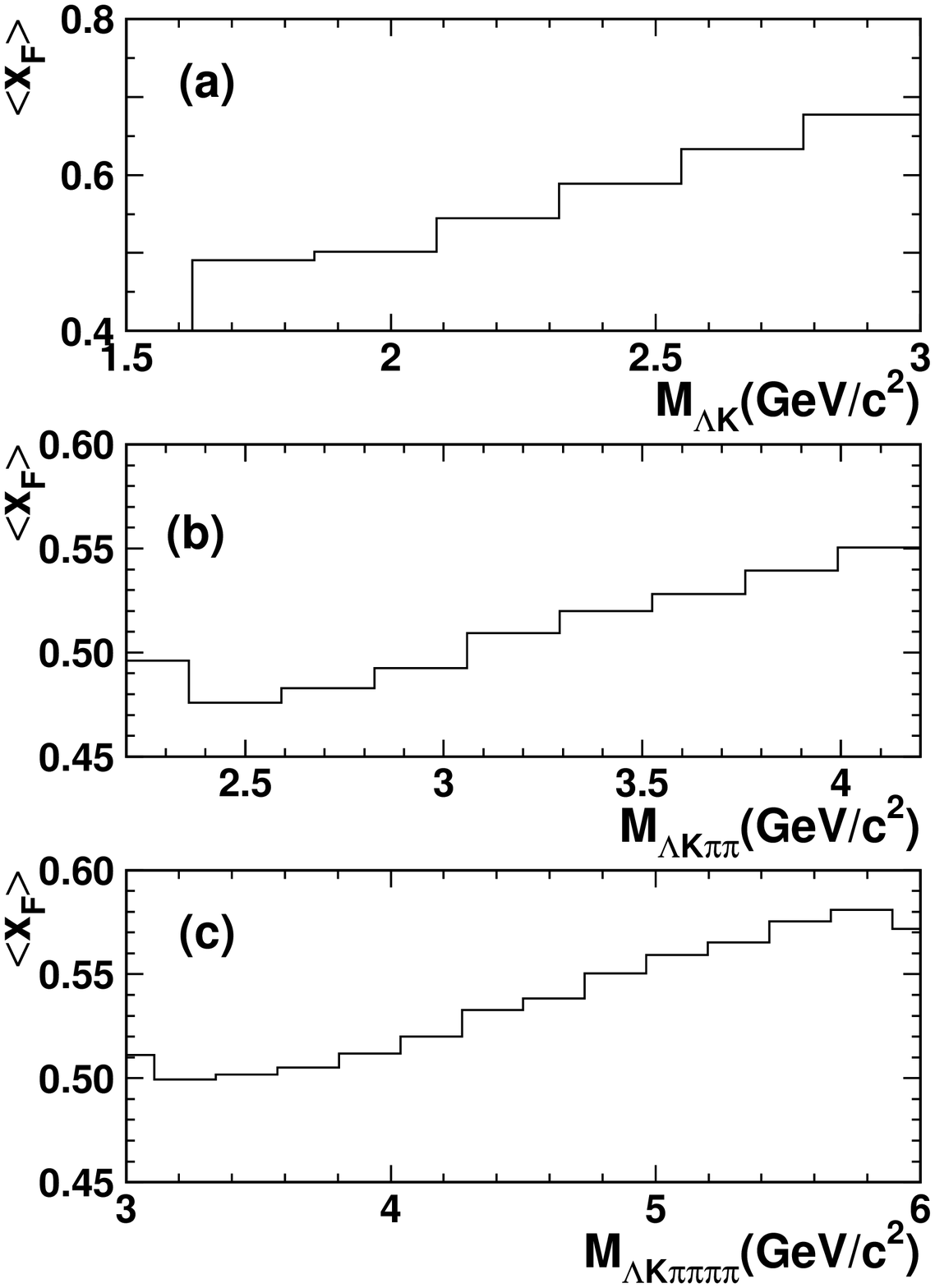,height=16cm}
\caption{Average values of $x_F$ 
of $\Lambda$ as a function 
of the invariant mass (a) $M_{\Lambda K}$,
(b) $M_{\Lambda K\pi\pi}$, 
or (c) $M_{\Lambda K\pi\pi\pi\pi}$ 
in $pp\to p\Lambda K^+\pi^+\pi^-\pi^+\pi^-$ at 27.5GeV/c.}
\end{figure}

\begin{figure}
\psfig{file=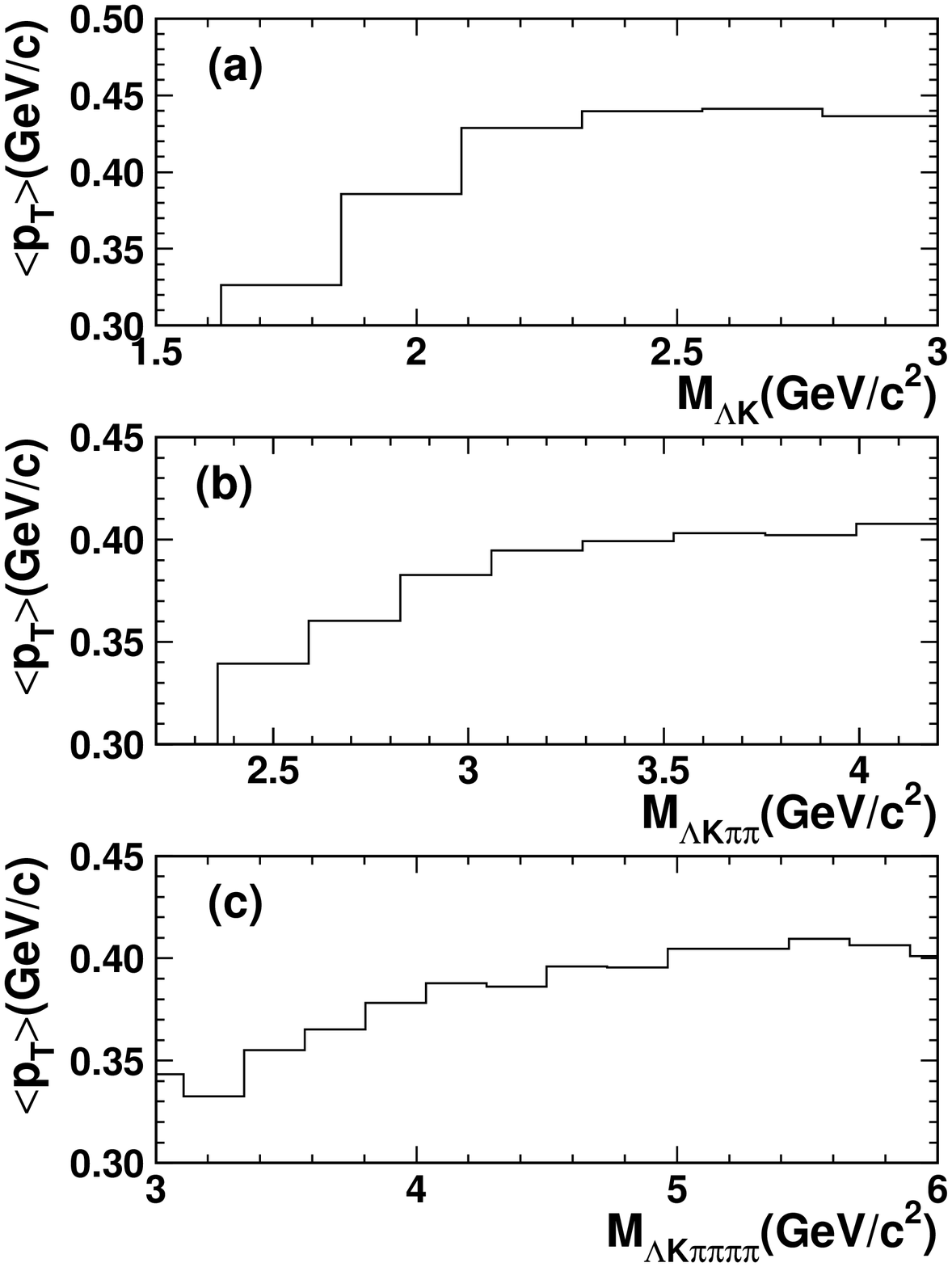,height=16cm}
\caption{Average values of $p_\perp$ 
of $\Lambda$ as a function 
of the invariant mass (a) $M_{\Lambda K}$,
(b) $M_{\Lambda K\pi\pi}$, 
or (c) $M_{\Lambda K\pi\pi\pi\pi}$ 
in $pp\to p\Lambda K^+\pi^+\pi^-\pi^+\pi^- $ at 27.5GeV/c.}
\end{figure} 

\begin{figure}
\psfig{file=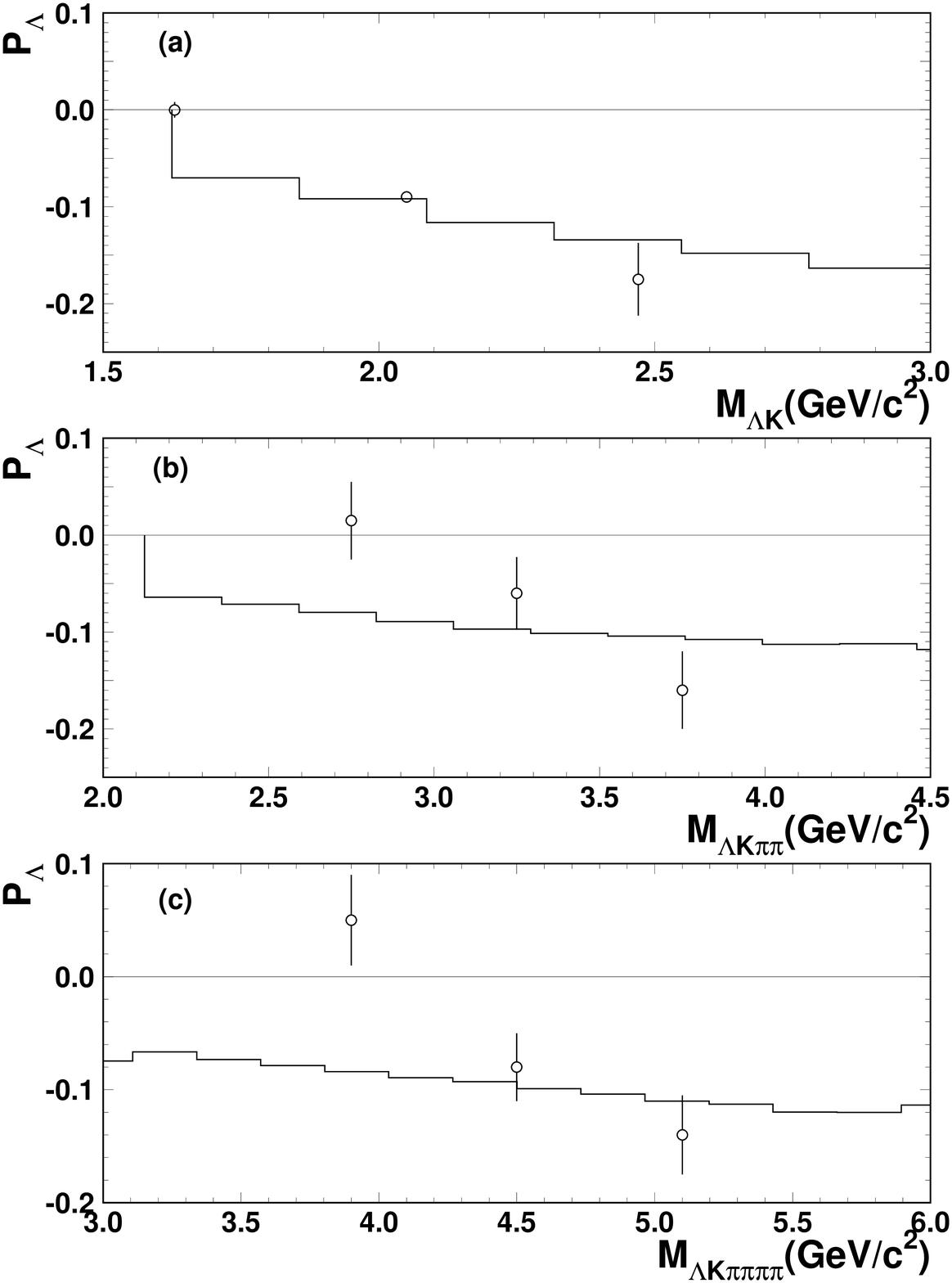,height=16cm}
\caption{$\Lambda$ polarization as a function 
of the invariant mass (a) $M_{\Lambda K}$,
(b) $M_{\Lambda K\pi\pi}$, 
or (c) $M_{\Lambda K\pi\pi\pi\pi}$ 
in $pp\to p\Lambda K^+\pi^+\pi^-\pi^+\pi^- $ at 27.5GeV/c.
The data are taken from \cite{Fel96}.}
\end{figure} 

\begin {thebibliography}{99}
\bibitem{foot} It should be mentioned in this connection that 
              similar measurements have been carried out 
              by R608 Collaboration at CERN 
              for the diffractive process $pp\to p\Lambda K^+$ at 
              $\sqrt{s}=63$ GeV (see [\ref{R608}]) and 
              similar increase of $|P_\Lambda|$ 
              with $M_{\Lambda K}$ has also been reported. 
              Clearly, the same kinematic effect as we discussed 
              above for $pp\to p\Lambda K^+\pi^+\pi^-\pi^+\pi^-$ 
              exists also here, i.e., 
              the $\langle x_F\rangle$ and $\langle p_\perp\rangle$ 
              for $\Lambda$ in this process increase 
              also with increasing $M_{\Lambda K}$. 
              Together with the $x_F$ dependence of $P_\Lambda$ 
              observed in $pp\to \Lambda X$, this effect leads 
              to an increase of $|P_\Lambda|$ with $M_{\Lambda K}$, 
              which agree qualitatively with the data.   
              It would be interesting also to extend our 
              quantitative estimations to this process 
              to see if the observed $M_{\Lambda K}$ dependence
              of $P_\Lambda$ can also be attributed to such effect. 
              This is unfortunately impossible at present, 
              since the $x_F$ values reached in the  
              R608 experiment can be as large as 0.96 which is far 
              beyond the $x_F$ region where we have data for $P_\Lambda$ 
              in $pp\to \Lambda X$. It is unclear whether the 
              parameterization of $P_\Lambda$ as a function 
              of $x_F$ and $p_\perp$ given in Eq.(2) is still valid 
              for such large $x_F$ and this parameterization plays 
              an essential role in our estimation.
              Hence, presently we can only conclude that, 
              there is definitely contribution  
              from the above-mentioned kinematic effect to the 
              $M_{\Lambda K}$ dependence of $P_\Lambda$ in that 
              diffractive process, 
              but we cannot say whether this contribution 
              can already account for 
              the increase observed in the experiment.
\end{thebibliography}
\end{document}